\def\ga{\mathrel{\mathchoice {\vcenter{\offinterlineskip\halign{\hfil
$\displaystyle##$\hfil\cr>\cr\sim\cr}}}
{\vcenter{\offinterlineskip\halign{\hfil$\textstyle##$\hfil\cr
>\cr\sim\cr}}}
{\vcenter{\offinterlineskip\halign{\hfil$\scriptstyle##$\hfil\cr
>\cr\sim\cr}}}
{\vcenter{\offinterlineskip\halign{\hfil$\scriptscriptstyle##$\hfil\cr
>\cr\sim\cr}}}}}

\documentclass[
   ,final            % use final for the camera ready runs
  ]
  {aipproc}

\usepackage{natbib}  
\layoutstyle{6x9}

\begin{document}

\title{Integral-Field Spectroscopy of SLACS Lenses}

%%%%%%%%%%%%%%%% TODO %%%%%%%%%%%%%%%%%%%%%%%%%%%%%%%%%%%%%%%%%%%%%%

\author{Oliver Czoske}{
  address={Kapteyn Astronomical Institute, P.O.~Box 800,
    9700~AV~Groningen, The~Netherlands}
}

\author{Matteo Barnab\`e}{
  address={Kapteyn Astronomical Institute, P.O.~Box 800,
    9700~AV~Groningen, The~Netherlands}
}

\author{L\'eon V. E. Koopmans}{
  address={Kapteyn Astronomical Institute, P.O.~Box 800,
    9700~AV~Groningen, The~Netherlands}
}

%%%\author{<author2>}{
%%%  address={<common address for author2 and author3>}
%%%}
%%%
%%%\author{<author3>}{
%%%  address={<common address for author2 and author3>}
%%%  ,altaddress={<author1 address>} % additional visiting address
%%%}

\begin{abstract}
  The combination of two-dimensional kinematics and gravitational lens
  modelling permits detailed reconstruction of the phase-space
  structure of early-type galaxies and sets constraints on the
  dark-matter distribution in their inner regions. We describe a
  project which combines integral-field spectroscopy from an ESO Large
  Programme using VIMOS on the VLT with deep HST ACS and NICMOS images
  to study a sample of 17~early-type lens galaxies at $z\approx 0.1 -
  0.3$, drawn from the Sloan Lens ACS survey (SLACS).
\end{abstract}

\maketitle

%%%%%%%%%%%%%%%%%%%%%%%%%%%%%%%%%%%%%%%%%%%%
%% MAINMATTER
%%%%%%%%%%%%%%%%%%%%%%%%%%%%%%%%%%%%%%%%%%%%

\section{Introduction}

Kinematic studies of early-type galaxies have a long history going
back to the pioneering work of Rudolph Minkowski in the 1950s
\citep{Burbidge1961a}. For a long time these studies were restricted
to zero- (aperture integrated) or at best one-dimensional spatial
resolution (using long slits). Over the last few years, new
instruments with integral-field spectroscopic capabilities have become
available on large telescopes that make it now possible to acquire
two-dimensional kinematic fields across the visible face of local
early-type galaxies.  The largest systematic observations of this type
to date have been performed using the SAURON instrument on the William
Herschel telescope, resulting in kinematic maps and dynamical models
of exquisite detail of a sample of 48~E~and S0 galaxies
\citep{Emsellem2004, Emsellem2007, Cappellari2007}. An even larger
follow-up project, ATLAS\textsuperscript{3D}, has recently been
started (see the contribution by P.~Serra in this volume).

Spatially resolved observations of galaxy kinematics become much more
difficult beyond the local universe, mainly due to the strong decrease
in the number of resolution elements available to cover the galaxies.
It is therefore useful to complement kinematic observations with other
types of information, such as gravitational lensing, which becomes
available for galaxies at redshifts $z\ga0.04$.

Like dynamics, lensing is sensitive to the total mass distribution
(visible and dark). It does, however, not require the matter to be in
dynamical equilibrium. Both dynamics and lensing individually show
degeneracies, preventing some interesting physical parameters from
being measured independently to good precision. A combined analysis
can break the degeneracies and thus provide information about the
galaxies that would not be available otherwise.

\section{Sloan Lens ACS Survey (SLACS)}

The Sloan Lens ACS Survey \citep{Bolton2006} has provided the largest
sample of gravitational lens systems to date. The survey uses the SDSS
luminous red galaxy sample and a quiescent subsample of the MAIN
sample as its parent samples. A galaxy is considered a lens candidate
if its SDSS spectrum shows emission lines at a higher redshift than
its absorption-line redshift. Candidates are subsequently imaged with
ACS or WFPC2 on HST to confirm the lens hypothesis and to provide the
high-resolution imaging which forms the basis for the construction of
detailed lens models. The full ACS sample currently comprises
89~confirmed lenses \citep{Bolton2008a}; WFPC2 observations of further
candidates are continuing.

Unlike many previous lens searches (e.g.~the CLASS survey,
\citep{Browne2003}), SLACS is a \emph{lens-selected} survey, i.e.~the
selection procedure guarantees the lenses to be bright and not
outshone by the gravitationally lensed background sources. This makes
SLACS a perfect sample for follow-up projects which combine the
lensing information with other types of observation, such as the
integral-field spectroscopy presented here.

\section{Observations}

\begin{figure}
  \includegraphics[height=.29\textheight]{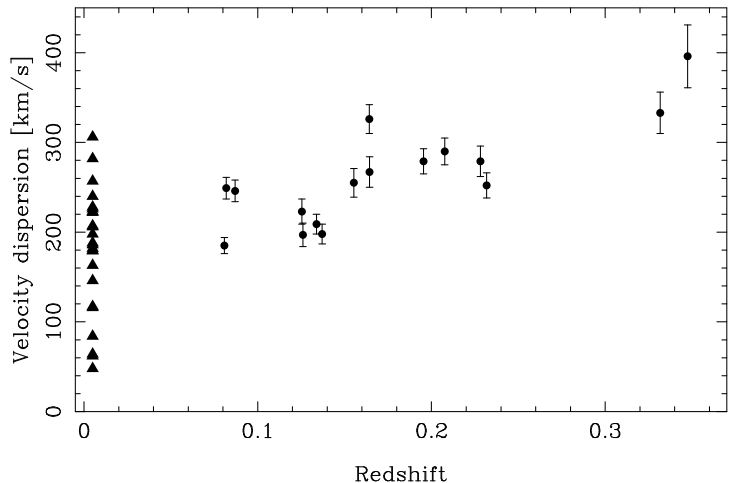}
  \caption{Distribution in redshift and velocity dispersion of the
    VIMOS-IFU (points with error bars, taken from \citep{Bolton2008a})
    and SAURON (triangles, taken from \citep{Emsellem2007}) samples.}
  \label{fig:sample-z-sigma}
\end{figure}

In an ESO Large Programme (177.B-0682), we have acquired
integral-field spectroscopy of 17~lenses from the SLACS sample using
the integral-field unit of VIMOS on the VLT UT3.
Fig.~\ref{fig:sample-z-sigma} shows the distribution of our sample in
redshift and velocity dispersion. Comparison to the SAURON sample
shows that our sample is roughly equivalent to the high-mass half of
the latter. However, our sample extends well beyond the local
universe, covering redshifts from $z=0.08$ to $0.35$ and thus
permitting studies of the possible evolution of the structure of
early-type galaxies over the last few Gyrs.

%%%%%%%%%%%%%%%%%%%%%%%%%%%%%%%%%%%%%%%%%%%%
%% Sample figure:
%%
%% The option [height=...] scales the picture to the given height,
%% without it it would be printed at its nominal size
%%%%%%%%%%%%%%%%%%%%%%%%%%%%%%%%%%%%%%%%%%%%

\section{Combined lensing and dynamics analysis}

The configuration of the gravitationally lensed source and the
kinematic data (velocity and velocity dispersion) are modelled
simultaneously in a fully self-consistent Bayesian framework. Full
details on the analysis method and code are given by
\citet{Barnabe-Koopmans2007}.

Briefly, the algorithm starts from a parametric, axially symmetric
model of the total gravitational potential, $\Phi(R, z, \eta_{k})$,
which in general depends on the parameters $\eta_{k}$ in a non-linear
way. For any given set of parameters, the lens reconstruction and the
dynamical reconstruction can be formulated as linear models; the
parameters to be reconstructed are the surface brightness distribution
in the source plane and the phase-space distribution,
respectively. The latter is built from two-integral components or
``TICs'' \citep{Verolme-deZeeuw2002}, which can be thought of as the
collection of all orbits with a given energy $E$ and angular
momentum~$L_{z}$.  The non-linear parameters of the gravitational
potential are then determined by maximising the Bayesian evidence,
permitting objective comparison of different types of models.

The result of the analysis are thus the best values for the non-linear
parameters $\eta_{k}$ and reconstructions of the lensed background
source and the phase space distribution function of the lens galaxy.

\section{First example: SDSS\,J2321$-$097}

\begin{figure}
  \centering
  \includegraphics[width=0.475\textwidth,angle=270]{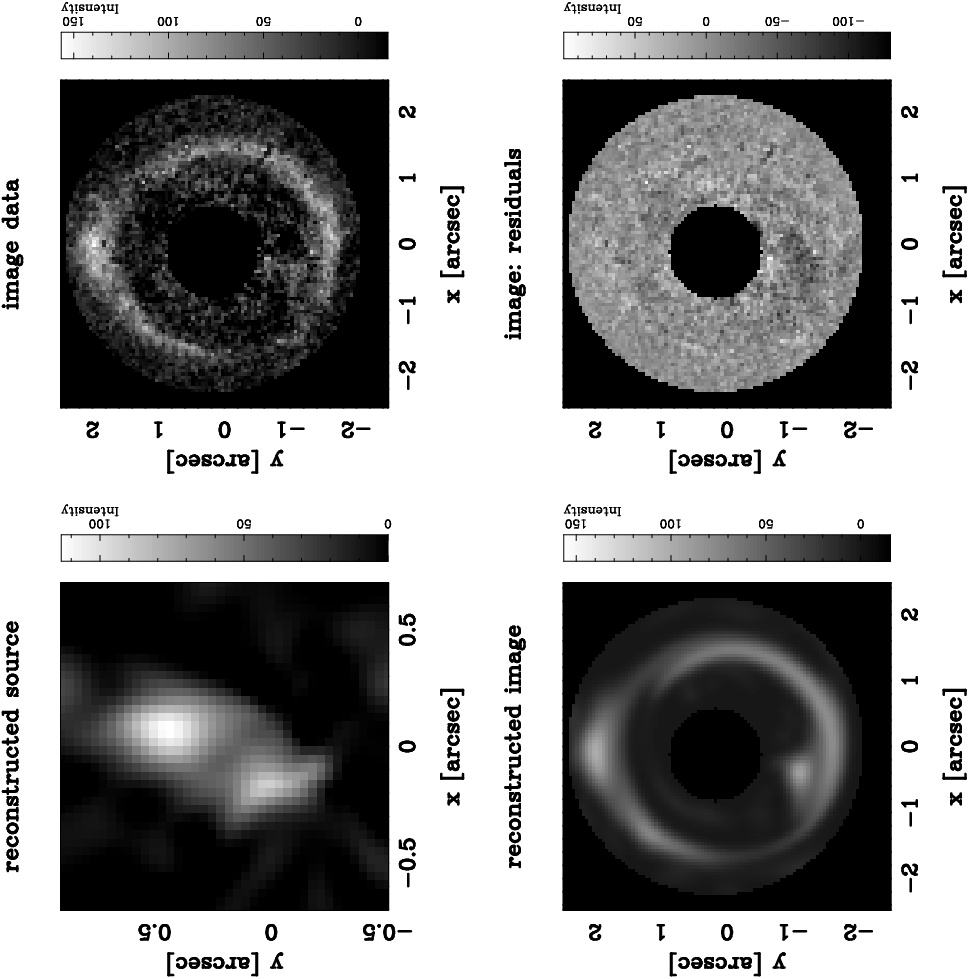}
  \hspace{0.05\textwidth}
  \includegraphics[width=0.475\textwidth,angle=270]{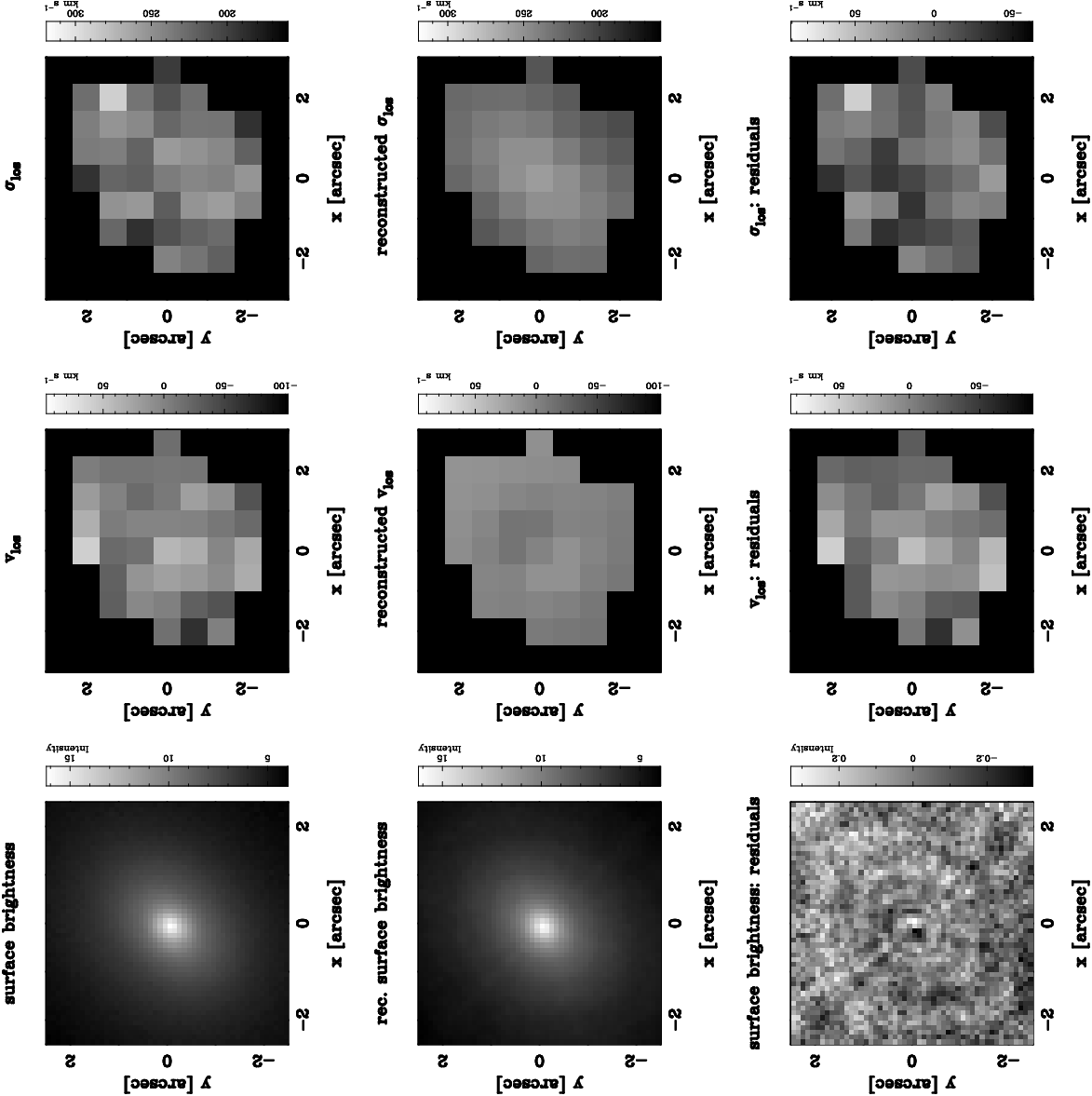}
  \caption{Lensing and dynamics reconstruction. The figures on the
    left show the observed gravitationally lensed source (after
    subtraction of the lens galaxy, top right), the reconstructed
    source in the source plane (top left) and in the lens plane
    (bottom left), as well as the residuals in the lens plane (bottom
    right). The figures on the right show the data in the top row:
    surface brightness of the lens galaxy, velocity $v(x, y)$ and
    velocity dispersion $\sigma(x, y)$. The middle row shows the
    reconstructions from the phase space distribution function, and
    the bottom row the residuals.}
  \label{fig:J2321-reconst}
\end{figure}

As an example, we show in Fig.~\ref{fig:J2321-reconst} the
two-dimensional reconstructions of the gravitationally lensed source
and the luminosity, velocity and velocity dispersion distributions in
the system SDSS\,J2321$-$097 at redshift $z=0.082$ \citep{Czoske2008}.

In most respects, SDSS\,J2321$-$097 is found to be a typical, massive
elliptical galaxy. Its velocity dispersion,
$\sigma\approx240\,\mathrm{km\,s^{-1}}$, would place it at the
high-mass end of the SAURON sample, and like the most massive SAURON
galaxies, SDSS\,J2321$-$097 is a pressure-supported slow rotator (as
defined by \citep{Emsellem2007}) with $v/\sigma \approx 0.1$.

In order to model the galaxy, a power-law mass density distribution
$\rho(r)\propto r^{\gamma}$ with axisymmetric ellipsoidal iso-density
surfaces was assumed.  The best-fitting logarithmic slope was found to
be $\gamma=-2.06^{+0.03}_{-0.06}$, i.e.~the density distribution is
fully consistent with being isothermal. This confirms results from
more simplistic analyses using the combination of lensing and
kinematic data (e.g.~\citep{Koopmans-Treu2002}). Disentangling the
contribution of dark and baryonic (i.e.~mostly stellar) matter to the
total density distribution requires further assumptions. In analogy to
the maximum disk approach in modelling rotation curves of spiral
galaxies we rescale the circularised stellar density profile (derived
from the observed surface brightness) such that it maximises the
contribution of the luminous component to the total density
profile. This `maximum bulge' approach provides a lower limit on
contribution of the non-luminous components of about 30~per cent
within the effective radius of the galaxy. This approach has been
tested on simulated $N$-body systems (which do not obey any
restrictive prescription of symmetry), showing that the correct dark
matter fraction is recovered within $\sim10\%$ \citep{Barnabe2008}.

\section{Outlook}

Apart from their use for reconstructing the mass structure of
early-type galaxies, a stellar population analysis of the IFU spectra
provides information on the age, metallicity and abundance ratios the
stellar component (cf.~\citep{Trager2008}). Also, the stellar
mass-to-light ratio can be estimated and used as further input towards
the final goal of disentangling the baryonic and dark contributions to
the total mass profile.

%%%%%%%%%%%%%%%%%%%%%%%%%%%%%%%%%%%%%%%%%%%%%%%%
%% BACKMATTER
%%%%%%%%%%%%%%%%%%%%%%%%%%%%%%%%%%%%%%%%%%%%%%%%

%% \begin{theacknowledgments}
%%   The work presented here has been done in collaboration with Matteo
%%   Barnab\`e, L\'eon Koopmans, Tommaso Treu, Adam Bolton and Scott
%%   Trager.
%% \end{theacknowledgments}



\bibliographystyle{aipproc}   % if natbib is available

\bibliography{articles}

\IfFileExists{\jobname.bbl}{}
 {\typeout{}
  \typeout{******************************************}
  \typeout{** Please run "bibtex \jobname" to obtain}
  \typeout{** the bibliography and then re-run LaTeX}
  \typeout{** twice to fix the references!}
  \typeout{******************************************}
  \typeout{}
 }

\end{document}